\documentclass[aps,pra,twocolumn,groupedaddress,showpacs,superscriptaddress]{revtex4-1}
\usepackage{amssymb,amsmath,graphicx,epstopdf,hyperref,amsthm}
\begin{document}
\interfootnotelinepenalty=10000

\title{Quantum nonlocal correlations are not dominated}

\author{Adrian Kent}
\email[]{A.P.A.Kent@damtp.cam.ac.uk }
\affiliation{Centre for Quantum Information and Foundations, DAMTP, Centre for
Mathematical Sciences,
University of Cambridge, Wilberforce Road, Cambridge, CB3 0WA, United Kingdom}
\affiliation{Perimeter Institute for Theoretical Physics, 31 Caroline Street
North, Waterloo, ON N2L 2Y5, Canada.}
\date{August 2013}

\begin{abstract}
We show that 
no probability distribution of spin measurement outcomes on pairs of
spin $1/2$ particles is unambiguously 
more nonlocal than the quantum correlations. 
That is, any distribution that 
produces a CHSH violation larger than the quantum violation 
for some axis choices also produces a smaller CHSH violation
for some other axis choices.  
In this sense, it is not possible for nature to 
be strictly more nonlocal than quantum theory allows.
\end{abstract}


\maketitle

Local hidden variable theories (LHVT) predict that the outcomes
of space-like separated measurements on particles should 
satisfy Bell inequalities \cite{Bell,CHSH69,BC90}.
Quantum theory predicts that Bell inequalities are violated
for suitable measurement choices on entangled particles. 

Bell inequalities and measures of nonlocality 
for two entangled qubits are
defined by considering experiments in which the 
corresponding particles 
are sent to two spacelike separated locations. 
One location is controlled by Alice, who performs 
measurement $A$; the other by Bob, who similarly performs $B$.
We focus here on the case where $A$ and $B$ are spin measurements about
given axes on spin-$\frac{1}{2}$ particles, in which
case Alice's and Bob's outcomes $a$ and $b$
are assigned values $a,b\in\lbrace1,-1\rbrace$,
corresponding to `spin up' or `spin down'.  
We define the correlation $C(A,B)$ as the average value of the product
of Alice's and Bob's outcomes in experiments where measurements $A$
and $B$ are chosen.

Consider for definiteness the EPR-Bohm experiment performed on
spin-$\frac{1}{2}$ particles in the singlet state
$\lvert\Psi^-\rangle=\frac{1}{\sqrt{2}}\bigl(\lvert \uparrow\rangle\lvert
\downarrow\rangle-\lvert
\downarrow\rangle\lvert \uparrow\rangle\bigr)$. 
As before Alice and Bob choose measurements
$A, B$ of their
particle spin projections along directions $\vec{a}_A$ and
$\vec{b}_B$, respectively.
In general, the vectors $\vec{a}_A$ and $\vec{b}_B$ can point along
any direction in 3-dimensional Euclidean space, and the sets of
their possible values define Bloch spheres $\mathbb{S}^2$.
The correlation predicted by quantum theory is $Q(\theta) = -\cos \theta$,
where $\cos \theta = \vec{a}_A \cdot \vec{b}_B$.
If Alice can choose between measurements $A$ and $A'$, and Bob
between $B$ and $B'$,  LHVT predict correlations that
satisfy the CHSH inequality \cite{CHSH69}: 
$$ I_2 = \bigl\lvert
C(A,B)+C(A,B' )+C(A' ,B)-C(A' , B' )\bigr\rvert \leq 2 \, .$$
On the other hand, sets of measurement axes can be found for which the quantum
correlations violate the CHSH inequality, $I_2^{QM} > 2$, up to the Cirel'son
\cite{C80} bound $I_2^{QM} \leq 2\sqrt{2}$.

Experiments compellingly confirm quantum theory and refute the
predictions of 
LHVT (e.g. \cite{ADR82,WJSWZ81,RKMSIMW01,MMMOM08,SBHGZ08,SBHGZ08}),
modulo possible loopholes (e.g. \cite{Pearle70,K05}) 
that arise from the difficulty in carrying
out theoretically ideal experiments.  
In this sense, quantum theory and nature are commonly said to 
exhibit {\it nonlocal} correlations.   
This is something of a misnomer, since there is a natural
locality principle respected by relativistic quantum
theory.  A more precise statement is that quantum
correlations are not {\it locally causal}, according to
Bell's definition \cite{Bellloccaus}.  
However, we will follow the common usage here, since it 
is adopted in most of the literature to which our work relates. 

Although nonlocal in the above sense,
quantum correlations do not allow superluminal signalling:
neither party's measurement choice affects the probability
distribution of the other's outcomes.
In an intriguing and celebrated paper \cite{PR94}, Popescu and Rohrlich
pointed out that quantum nonlocal correlations are not characterised
by the no-signalling condition alone. 
They illustrated this using what they called a ``superquantum''
correlation function $E$ for spin measurements about given axes, 
defined by a probability distribution whose marginals are uniform
for any spin measurement by either party, conditioned on any
measurement choice of the other party.  
The function $E$ depends only on the relative angle $\theta$ between axes
and has the form 
\begin{itemize}
\item  $ E ( \theta ) = 1 $ for $ 0 \leq \theta \leq \pi /4 $.
\item $E ( \theta )$ decreases monotonically and smoothly from $1$
to $0$ as $\theta$ increases from $\pi/4$ to $\pi/2$.
\item $E(\pi - \theta) = - E( \theta )$ for $\pi/2 \leq \theta \leq
  \pi$.
\end{itemize}

For coplanar axes $\vec{a'}, \vec{b}, \vec{a}, \vec{b'}$ 
separated by successive $\pi/4$
rotations, this gives the algebraically maximal CHSH expression 
\begin{eqnarray}
E(\vec{a},\vec{b}) + E(\vec{a}',\vec{b}) + E(\vec{a},\vec{b}') - 
E(\vec{a}',\vec{b}') &=& \nonumber \\ 3 E( \pi/4 ) - E (3 \pi /4 ) 
&=& 
4  \, . 
\end{eqnarray} 
This violates the Cirel'son bound for quantum correlations, 
but still follows from a non-signalling probability distribution.  

Taking this CHSH expression as a measure of nonlocality, 
Popescu and Rohrlich went on to ask why quantum theory
is not more non-local and whether stronger forms
of nonlocality might be found in nature. 

This raises a question: {\it are} the Popescu-Rohrlich
correlations, or any other hypothetical sets of non-signalling 
correlations, unambiguously more non-local than quantum correlations?  
To even make sense of the question, one has to accept the premise 
that there is at least a partial ordering of the non-locality
of correlations, which is reflected by the degree of
violation of Bell inequalities.   
Then one needs to decide which Bell inequalities to
consider.  It is far from obvious that there is a
natural way to do this.  Even for the CHSH inequality,
there are infinitely many possible axis choices to
consider.  Moreover, the CHSH inequality is only 
one of an infinite number of Bell inequalities defining  
different possible tests of quantum non-locality.

Of course, one possible candidate measure of non-locality is
the maximum CHSH violation that a set of correlations gives,
for any set of axis choices, and the Popescu-Rohrlich correlations
are more non-local than quantum theory by this measure. 
But the maximum violation is not the only possible measure, and it 
is arguable whether it is the most natural. 

In this paper, we answer the question above in the negative:
neither the Popescu-Rohrlich correlations nor any others are
{\it unambiguously} more non-local than the quantum singlet correlations. 

More precisely, we consider correlation functions  $C (\theta )$
defined by hypothetical probability distributions for spin measurements
of two particles about randomly chosen axes separated by angle
$\theta$.   Here $C( \theta )$ is the 
correlation averaged over all pairs of axes
separated by $\theta$: the actual probability distributions
may depend on the axis choices as well as their angular separation.   
Since spin measurement outcomes correspond to a positive
or negative axis vector, we take $C( \pi - \theta ) = -
C(\theta )$.   We also make the physically motivated assumption
that $C ( \theta )$ depends continuously on $\theta$.  

We then show that if any such
$C(\theta )$ produces a larger violation of some
Bell inequality than the singlet quantum correlations do,
then $C(\theta )$ must produce a smaller violation (or none) of 
some other Bell inequality also violated by the singlet quantum
correlations.
In other words, any correlations that, by a measure analogous to that 
used by Popescu-Rohrlich, are ``more nonlocal'' than the singlet are also,
by another such measure, ``less nonlocal''. 
This is true whether or not the 
correlations arise from a non-signalling probability distribution, so
long as the underlying theory defines correlation functions $C( \theta
)$ that depend only on the angular separation $\theta$ and not on the
details of how the ensemble of measurements with given $\theta$ is 
produced.   
   
We use the following CHSH inequalities.   The CHSH expression
for quantum spin measurements about a randomly chosen set of
coplanar axes $\vec{a}' , \vec{b} , \vec{a} , \vec{b}'$ 
separated by angles $\theta , \pi/2 - \theta, \theta$ respectively,
gives
\begin{eqnarray}
I^{CHSH1}_{QM} ( \theta ) &=& | 2 C_{QM} ( \theta ) + 2 C_{QM} ( \pi/2 -
\theta ) | \nonumber \\ 
&=& 2 \cos \theta + 2 \cos ( \pi/2 - \theta ) \nonumber \\ &>& 2 \, , 
\end{eqnarray}
for $0 < \theta < \pi/2$, violating the CHSH inequality
\begin{equation}
I^{CHSH1} \leq 2 \,. 
\end{equation}

The CHSH expression for quantum spin measurements about a randomly
chosen set of coplanar axes 
$\vec{a}' , \vec{b} , \vec{a} , \vec{b}' $ separated by
angles $\theta/3 , \theta/3, \theta/3$ respectively gives
\begin{equation}
I^{CHSH2}_{QM} ( \theta ) = | 3 \cos (\theta /3 ) - \cos  \theta | >2
, 
\end{equation} 
for $0 < \theta < \pi/2$, violating the CHSH inequality 
\begin{equation}
I^{CHSH2} \leq 2 \, .   
\end{equation}

Any correlation function $C( \theta )$ that is ``at least as
non-local'' as quantum theory according to these inequalities
must thus satisfy
\begin{equation} \label{bcnonlocal}
| 2 C ( \theta ) + 2 C ( \pi/2 - \theta ) | \geq I^{CHSH1}_{QM} ( \theta
)> 2 \, . 
\end{equation}
and
\begin{equation} \label{chshnonlocal}
| 3 C( \theta/ 3) - C (  \theta ) | \geq I^{CHSH2}_{QM} ( \theta ) >2 
 \, . 
\end{equation}
for $0 < \theta < \pi/2$. 

Consider a hypothetical $C ( \theta )$ with these properties.

Note first that $C( \theta )$ must have the same sign throughout
the range $0 < \theta < \pi/2$.  If not, then by continuity 
$C ( \theta_0 ) = 0 $ for some $\theta_0$ in the range, and then
(\ref{bcnonlocal}) fails at $\theta= \theta_0$.

Now consider the case in which $C ( \theta ) < 0 $ for all
$\theta$ in the range.  Suppose that for some $\theta$ we
have $0 > C( \theta ) > - \cos \theta$.   It follows from
(\ref{bcnonlocal}) that 
\begin{equation}
C(\pi/2 - \theta ) < - \cos ( \pi/2 -
\theta ) \, .
\end{equation}
Hence either $C( \theta ) = - \cos \theta$ for all
$\theta$ in the range, in which case $C$ is the
quantum correlation function, or $C( \theta ) < - \cos \theta$ for at
least one value of $\theta$ in the range. 

But now, if $C ( \theta_1 ) = - \cos \theta_1 - \delta$, for some $\delta
>0$ and some $\theta_1$ in the range,
then applying (\ref{chshnonlocal}) iteratively gives
\begin{equation} \label{smallboundone}
- \cos ( \theta_1 3^{-n} ) - C ( \theta_1 3^{-n} ) \geq \delta 3^{-n} \, .
\end{equation}
However, since 
\begin{equation}
- \cos ( \theta_1 3^{-n} ) \leq - 1 + \theta_1^2  2^{-1} 3^{-2n} 
\end{equation} 
for large $n$, and $C (\theta ) \geq -1$ for all $\theta$,
we have 
\begin{equation} \label{smallboundtwo}
- \cos ( \theta_1 3^{-n} ) - C( \theta_1 3^{-n} ) \leq \theta_1^2 2^{-1}
3^{-2n}
\end{equation}
for large $n$, contradicting (\ref{smallboundone}).
Hence $C( \theta ) = - \cos \theta$ for all $\theta$ in the range.

Similarly, if $C( \theta ) > 0 $ for all $\theta$ in the range,
we find $C(\theta) = \cos \theta$.  This is the correlation obtained
from quantum theory if one party reverses their 
measurement outcome.   As these are the only possibilities, we see 
that no super-quantum correlation functions -- in the sense
we have defined -- exist.

\section{Discussion}

Following Popescu and Rohrlich, we 
have focussed on hypothetical generalisations of
the quantum correlations of a pair of entangled qubits, 
although it would be interesting to extend the discussion further to higher 
dimensions and to multipartite states.   
We have shown that no theory
can produce spin measurement correlations for pairs of spin $1/2$
particles that dominate quantum nonlocal correlations, in the sense
that they are at least as nonlocal by every measure and more nonlocal
by at least one measure.  

This observation suggests another way of looking at 
Popescu and Rohrlich's intriguing observations. 
As other recent results \cite{kent2013sphere} also suggest, 
degrees of quantum nonlocality can only be properly compared  
when we consider the full range of possible spin
measurements allowed by physics, rather than restricting attention
to measures of nonlocality associated with particular
finite sets of measurement axis choices.   
However, if we look at the full range of spin measurements, we see 
that neither the correlations that Popescu and Rohrlich consider nor
any other possible set of correlations are 
unambiguously more strongly non-local than those of quantum theory.

Quantum correlations for entangled spin $1/2$ particles
take the precise form they do because
they reflect the interrelation between the Bloch sphere representations
of the local spin rotation group $SU(2)$ and the 
local spatial rotation group
$SO(3)$.   One reason to think that Popescu and Rohrlich's ``superquantum''
correlations may not arise in nature
is that they do not arise naturally from local 
physical symmetries in this way.  

As Popescu and Rohrlich originally framed the question, 
the maximum quantum CHSH value of $2 \sqrt{2}$ sits interestingly
between the classical value of $2$ and the value 
of $4$ attainable by non-signalling correlations.   From this 
perspective, the number $2 \sqrt{2}$ seems a puzzle in need
of explanation. 

One intriguing line of thought suggests
that the explanation is to be found in the relation between
physics and the information capacity of messages.  Super-quantum
correlations that violate the  Cirel'son bound also violate the
principle of information causality \cite{ic}.   If this is a 
fundamental principle of nature, then the puzzle is solved.  

Whether the laws of nature {\it are} fundamentally information-theoretic
in this or other respects, is though, presently uncertain.   
We are far from fully understanding the fundamental principles 
underlying physics, and different ways of looking
at deep unresolved questions can seem to strongly suggest different 
answers.   With that important caveat noted, 
our results suggest another possible perspective.   
The maximal quantum CHSH violation of
$2 \sqrt{2} = 4 \cos ( \pi / 4)$ may suggest a puzzle when
considered in isolation.  However, the full set of CHSH violations
given by the singlet correlation function $ - \cos ( \theta )$
do not seem analogously puzzling, since they are not dominated by any
other correlations.   

In this significant sense, the Popescu-Rohrlich correlations are
not actually ``superquantum''.   It 
is true that the two-input two-output ``nonlocal boxes''
given by specific measurement axes are extreme points in the space
of non-signalling correlations, while the quantum correlations for
any state and any pairs of axis choices are not.   
However, if we look at the full 
set of correlations for all angles $\theta$, the Popescu-Rohrlich
correlations are no longer distinguished from quantum singlet correlations
by this criterion.   
From this perspective, then, there is perhaps 
really no fundamental puzzle about why 
Popescu-Rohrlich or other purportedly ``super-quantum'' correlations
are not used by nature, since there seems no 
sufficiently strong theoretical reason to think they characterise 
singularly physically interesting generalizations of quantum 
theory in our space-time.

\begin{acknowledgments}

I thank Dami\'an Pital\'ua-Garc\'ia
for very helpful comments and discussions. 
This work was partially
supported by a grant from the John Templeton Foundation and by
Perimeter Institute for Theoretical Physics. Research at Perimeter
Institute is supported by the Government of Canada through Industry
Canada and by the Province of Ontario through the Ministry of
Research and Innovation.

\end{acknowledgments}

\bibliography{qcorrbiblio}
\end{document}